\pdfoutput=1 
\documentclass[aps,twocolumn,prl,showpacs,amsmath,amssymb,10pt]{revtex4-1}
\usepackage{graphicx}
\usepackage{scrextend}
\usepackage{manfnt,pifont}
\usepackage{hyperref}
\usepackage{url}

\usepackage{url}

\begin{document}



\title{Double Copy Relation for AdS}

\author{Xinan Zhou}

\affiliation{Princeton Center for Theoretical Science, Princeton University, Princeton, NJ 08544, USA}


\begin{abstract}
\noindent We present a double copy relation in AdS$_5$ which relates tree-level four-point amplitudes of supergravity, super Yang-Mills and bi-adjoint scalars. 
\end{abstract}

\maketitle

\section{Introduction}
Scattering amplitudes in flat space exhibit surprising properties which encode deep lessons for quantum field theories and gravity. While we believe many curved-spacetime generalizations exist, explicit realizations are far from obvious to find. Recently, there has been a lot of activity trying to extend two remarkable flat-space properties, color-kinematic duality \cite{Bern:2008qj} and the double copy relation \cite{Bern:2010ue}, to the simplest curved background -- AdS space \cite{Farrow:2018yni,Lipstein:2019mpu,Armstrong:2020woi,Albayrak:2020fyp,Alday:2021odx} \footnote{See, {\it e.g.},  \cite{Adamo:2017nia,Adamo:2018mpq} for progress in other curved backgrounds.}. The flat-space relations relate gauge theory and gravity amplitudes, and have numerous applications in modern amplitude research \footnote{See \cite{Bern:2019prr} for a recent comprehensive review.}. Since AdS/CFT maps AdS amplitudes to CFT correlators, generalizations to AdS are especially interesting. While color-kinematic duality has been observed for four points \cite{Armstrong:2020woi,Albayrak:2020fyp,Alday:2021odx}, AdS double copy so far has only worked for three-point functions \cite{Farrow:2018yni,Lipstein:2019mpu}. In fact, it was not clear if the flat-space relation has to be drastically modified at higher points. In this paper, we present an AdS generalization which realizes the double copy construction in four-point amplitudes for the first time. We relate tree-level amplitudes in AdS$_5$$\times$S$^5$ IIB supergravity, AdS$_5$$\times$S$^3$ SYM, and non-supersymmetric AdS$_5$$\times$S$^1$ bi-adjoint scalars, in a simple way that mirrors the flat space relation. Moreover, our AdS relation works for {\it all} amplitudes in these theories, applying to massless and massive particles alike.

We will use the Mellin representation for CFT correlators \cite{Mack:2009mi,Penedones:2010ue}. AdS amplitudes become Mellin amplitudes and enjoy simple analytic structure resembling the flat-space one. Tree-level Mellin amplitudes of AdS supergravity and super gauge theories in various spacetime dimensions were systematically studied in \cite{Rastelli:2016nze,Rastelli:2017udc,Rastelli:2019gtj,Giusto:2019pxc,Giusto:2020neo,Alday:2020lbp,Alday:2020dtb,Alday:2021odx,Wen:2021lio}, and a Mellin color-kinematic relation similar to the flat-space one was pointed out in \cite{Alday:2021odx}. Unfortunately, applying the flat-space double copy prescription led to no sensible amplitudes. In this paper, we revisit these results. We will focus on AdS$_5$ and take advantage of supersymmetry, which allows us to reduce the Mellin amplitudes to simpler {\it reduced} Mellin amplitudes. We find that it is in these reduced objects that color-kinematic duality and double copy relation are naturally realized.

Schematically, we will write the reduced amplitude of AdS$_5$ super gluons with $\mathcal{N}=2$ superconformal symmetry as a finite sum labelled by integers $i$, $j$
\begin{equation}
\nonumber \widetilde{\mathcal{M}}\sim \sum_{i,j}\frac{\mathtt{n}^{i,j}_s\mathtt{c}_s}{s-s_{i,j}}+\frac{\mathtt{n}^{i,j}_t\mathtt{c}_t}{t-t_{i,j}}+\frac{\mathtt{n}^{i,j}_u\mathtt{c}_u}{u-u_{i,j}}\;,
\end{equation}
where the number of terms is determined by the external masses. $\mathtt{c}_{s,t,u}$ are standard color factors satisfying $\mathtt{c}_s+\mathtt{c}_t+\mathtt{c}_u=0$. The kinematic factors $\mathtt{n}^{i,j}_{s,t,u}$ turn out to obey the same relation $\mathtt{n}^{i,j}_s+\mathtt{n}^{i,j}_t+\mathtt{n}^{i,j}_u=0$, giving rise to an AdS color-kinematic duality. Replacing $\mathtt{c}_{s,t,u}$ with $\mathtt{n}^{i,j}_{s,t,u}$, we recover precisely super graviton reduced amplitudes of AdS$_5$$\times$S$^5$ IIB supergravity \cite{Rastelli:2016nze,Rastelli:2017udc}. On the other hand, replacing $\mathtt{n}^{i,j}_{s,t,u}$ by $\mathtt{c}_{s,t,u}$ leads to Mellin amplitudes of conformally coupled bi-adjoint scalars on AdS$_5$$\times$S$^1$, which were not studied in the literature. We will prove it by direct calculation. The AdS$_5$ double copy relation presented here relates theories with varying $\mathcal{N}=0,2,4$ superconformal symmetry. However, it also implies that purely bosonic theories of Einstein gravity, YM, and bi-adjoint scalars on AdS$_5$ should be related by double copy, as we will briefly discuss at the end.  

\section{Four-point correlators}

\noindent{\bf No supersymmetry.} Let us start with the non-supersymmetric case. We consider the correlator of four scalar operators $\mathcal{O}_{k_i}$ with conformal dimensions $k_i$ \footnote{In this paper, the two-point functions of all $\mathcal{O}_k$ are unit normalized. }
\begin{equation}\label{G4pt}
G_{k_1k_2k_3k_4}=\langle \mathcal{O}_{k_1} \mathcal{O}_{k_2} \mathcal{O}_{k_3} \mathcal{O}_{k_4} \rangle\;.
\end{equation}
In Mellin space correlators are represented as \cite{Mack:2009mi,Penedones:2010ue}
\begin{equation}\label{MellinG}
G_{k_1k_2k_3k_4}\!=\!\!\!\!\int_{-i\infty}^{i\infty}\!\!\![dsdt]{\rm K}(x_{ij}^2;s,t,u) \mathcal{M}_{k_1k_2k_3k_4} \Gamma_{\{k_i\}}(s,t,u)
\end{equation}
where $[dsdt]=\frac{dsdt}{(4\pi i)^2}$, and ${\rm K}(x_{ij}^2;s,t,u)$ is a factor containing all spacetime dependence
\begin{equation}
\begin{split}
\nonumber {\rm K}(x_{ij}^2;s,t,u)=&(x_{12}^2)^{\frac{s-k_1-k_2}{2}}(x_{34}^2)^{\frac{s-k_3-k_4}{2}}(x_{14}^2)^{\frac{t-k_1-k_4}{2}}\\
\times &(x_{23}^2)^{\frac{t-k_2-k_3}{2}}(x_{13}^2)^{\frac{u-k_1-k_3}{2}}(x_{24}^2)^{\frac{u-k_2-k_4}{2}}\;.
\end{split}
\end{equation}
Here $x_{ij}=x_i-x_j$, and $s$, $t$, $u$ are Mandelstam variables satisfying $s+t+u=\sum_{i=1}^4 k_i\equiv \Sigma$ \footnote{Note that only two of the three Mandelstam variables are independent. While rewriting a function of $s$, $t$ in terms of $s$, $t$, $u$ is not unique, there is no ambiguity to define expressions using all three variables.}.  
 We have also extracted a factor of Gamma functions 
\begin{equation}
\begin{split}
\Gamma_{\{k_i\}}(s,t,u)=&\Gamma[\tfrac{k_1+k_2-s}{2}]\Gamma[\tfrac{k_3+k_4-s}{2}]\Gamma[\tfrac{k_1+k_4-t}{2}]\\
\times &\Gamma[\tfrac{k_2+k_3-t}{2}]\Gamma[\tfrac{k_1+k_3-u}{2}]\Gamma[\tfrac{k_2+k_4-u}{2}]\;.
\end{split}
\end{equation}
which captures the contribution of double-trace operators universally present in the holographic limit \cite{Penedones:2010ue} . All  dynamic information is contained in $\mathcal{M}_{k_1k_2k_3k_4}$,  known as the {\it Mellin amplitude}. The four-point function $G_{k_1k_2k_3k_4}$ obeys Bose symmetry which permutes operators. Bose symmetry acts on the Mellin amplitude by interchanging $k_i$, as well as permuting the Mandelstam variables $s$, $t$, $u$ -- in the same way it acts on a flat-space amplitude.

\vspace{0.2cm}

\noindent{\bf $\mathcal{N}=2$ superconformal symmetry.} We now consider CFTs with $\mathcal{N}=2$ superconformal symmetry, focusing on the $\frac{1}{2}$-BPS operators. These operators are of the form $\mathcal{O}_{k}^{a_1\ldots a_k}$ where $a_i=1,2$ are indices of the R-symmetry group $SU(2)_R$ \footnote{There is also a $U(1)_r$ R-symmetry in the $\mathcal{N}=2$ superconformal algebra. But the $\frac{1}{2}$-BPS operators are not charged under $U(1)_r$.}. The operator $\mathcal{O}_{k}^{a_1\ldots a_k}$ transforms in the spin $j_R=\frac{k}{2}$ representation of $SU(2)_R$, and has conformal dimensions $k=2,3,\ldots$. To conveniently keep track of the  $SU(2)_R$ indices, we contract them with auxiliary two-component spinors $v^a$
\begin{equation}
\mathcal{O}_k(x,v)=\mathcal{O}_{k}^{a_1\ldots a_k}v^{b_1}\ldots v^{b_k}\epsilon_{a_1b_1}\ldots \epsilon_{a_kb_k}\;.
\end{equation}
We then consider their four-point functions (\ref{G4pt}), and define the Mellin amplitude $\mathcal{M}_{k_1k_2k_3k_4}^{\mathcal{N}=2}$ via (\ref{MellinG}). 

The $\mathcal{N}=2$ superconformal symmetry imposes extra constraints on the form of correlators via the superconformal Ward identities \cite{Nirschl:2004pa}. Solving them leads to
\begin{equation}\label{WardNeq2}
G_{k_1k_2k_3k_4}^{\mathcal{N}=2}=G_{0,k_1k_2k_3k_4}^{\mathcal{N}=2}+{\rm R}^{(2)}H_{k_1k_2k_3k_4}^{\mathcal{N}=2}
\end{equation}
where $G_{0,k_1k_2k_3k_4}^{\mathcal{N}=2}$ is the protected part of the correlator independent of marginal couplings. The factor ${\rm R}^{(2)}$ is crossing symmetric, and is fixed by superconformal symmetry to be
\begin{equation}\label{R2}
{\rm R}^{(2)}=(v_1\cdot v_2)^2(v_3\cdot v_4)^2 x_{13}^2x_{24}^2(1-z\alpha)(1-\bar{z}\alpha)\;.
\end{equation}
Here $v_i\cdot v_j=v_i^a v_j^b\epsilon_{ab}$, $\alpha=\frac{(v_1\cdot v_3)(v_2\cdot v_4)}{(v_1\cdot v_2)(v_3\cdot v_4)}$ is the $SU(2)_R$ cross ratio, and $z$, $\bar{z}$ are conformal cross ratios given by
\begin{equation}\label{cfcrossratios}
z\bar{z}=\frac{x_{12}^2x_{34}^2}{x_{13}^2x_{24}^2}=U\;,\quad (1-z)(1-\bar{z})=\frac{x_{14}^2x_{23}^2}{x_{13}^2x_{24}^2}=V\;.
\end{equation}
All the dynamical information is contained in the simpler {\it reduced} correlator $H_{k_1k_2k_3k_4}^{\mathcal{N}=2}$, which can be viewed as a correlator of operators with {\it shifted} conformal dimensions $k_i+1$ and shifted $SU(2)_R$ spins $\frac{k_i}{2}-1$. 

In the regime to be considered, corresponding to AdS tree level, the reduced correlator in fact captures {\it all} the information. To make it more precise, let us define a {\it reduced} Mellin amplitude via the reduced correlator
\begin{equation}
\nonumber H_{k_1k_2k_3k_4}^{\mathcal{N}=2}\!=\!\!\!\int_{-i\infty}^{i\infty}\!\!\![dsdt]{\rm K}(x_{ij}^2;s,t,\tilde{u}) \widetilde{\mathcal{M}}_{k_1k_2k_3k_4}^{\mathcal{N}=2} \Gamma_{\{k_i\}}(s,t,\tilde{u})\,.
\end{equation}
Note that it is important to shift the $u$ variable to $\tilde{u}=u-2$ so that $s+t+\tilde{u}=\Sigma-2$. The shift is to compensate the nonzero weights of the factor ${\rm R}^{(2)}$ under conformal transformations. As a consequence, Bose symmetry acts differently in the full and reduced Mellin amplitudes, as
\begin{equation}\label{reducedpermu}
\begin{split}
&\mathcal{M}_{k_1k_2k_3k_4}^{\mathcal{N}=2}\;:\quad \text{permuting }s\,,t\,,u\;,\\
&\widetilde{\mathcal{M}}_{k_1k_2k_3k_4}^{\mathcal{N}=2}\;:\quad \text{permuting }s\,,t\,,\tilde{u}\;.
\end{split}
\end{equation}
In the tree-level regime the protected part $G_{0,k_1k_2k_3k_4}^{\mathcal{N}=2}$ does {\it not} contribute to the Mellin amplitude \cite{Alday:2021odx}. Rather it is generated by a contour pinching mechanism described in \cite{Rastelli:2017udc}. Therefore full amplitudes are {\it completely} determined by reduced amplitudes, with the precise relation given by translating both sides of (\ref{WardNeq2}) into Mellin space
\begin{equation}
\mathcal{M}_{k_1k_2k_3k_4}^{\mathcal{N}=2}=\mathbb{R}^{(2)}\circ \widetilde{\mathcal{M}}_{k_1k_2k_3k_4}^{\mathcal{N}=2}\;.
\end{equation}
The factor ${\rm R}^{(2)}$ now becomes a difference operator $\mathbb{R}^{(2)}$ \cite{Alday:2021odx}. To obtain it, we interpret each monomial $U^mV^n$ in 
\begin{equation}
\frac{{\rm R}^{(2)}}{(v_1\cdot v_2)^2(v_3\cdot v_4)^2 x_{13}^2x_{24}^2}
\end{equation}
as a difference operator $U^mV^n\to \mathbb{O}_{m,n}^{(2)}$ which acts on functions $f(s,t)$ according to 
\begin{equation}\label{Oaction}
\begin{split}
\mathbb{O}_{m,n}^{(\mathcal{N})}\circ f(s,t)=&\tfrac{\Gamma_{\{k_i\}}(s-2m,t-2n,\Sigma-\mathcal{N}-s-t+2m+2n)}{\Gamma_{\{k_i\}}(s,t,\Sigma-s-t)}\\
&\times f(s-2m,t-2n)
\end{split}
\end{equation}

\vspace{0.2cm}

\noindent{\bf $\mathcal{N}=4$ superconformal symmetry.} The kinematics of $\mathcal{N}=4$ is similar. The $\frac{1}{2}$-BPS operator, labelled by an integer $k=2,3,\ldots$, transforms in the rank-$k$ symmetric traceless representation of the $SO(6)_R$ R-symmetry group, and has dimension $k$. We keep track of the R-symmetry indices by using null $SO(6)$ vectors $t^r$ \footnote{These null vectors automatically project the indices to the symmetric traceless representation.}
\begin{equation}
O_k(x,t)=O^{r_1\ldots r_k}(x)t^{r_1}\ldots t^{r_k}\;,\quad r_i=1,\ldots,6
\end{equation}
where $t\cdot t=0$. The $\mathcal{N}=4$ superconformal symmetry dictates that the four-point function is of the form \cite{Eden:2000bk,Nirschl:2004pa}
\begin{equation}\label{WardNeq4}
G_{k_1k_2k_3k_4}^{\mathcal{N}=4}=G_{0,k_1k_2k_3k_4}^{\mathcal{N}=4}+{\rm R}^{(4)}H_{k_1k_2k_3k_4}^{\mathcal{N}=4}
\end{equation}
where $G_{0,k_1k_2k_3k_4}^{\mathcal{N}=4}$ is the protected part, and $H_{k_1k_2k_3k_4}^{\mathcal{N}=4}$ is the {\it reduced} correlator. Note that the reduced correlator also has shifted quantum numbers, with dimensions $k_i+2$ and $SO(6)$ spin $k_i-2$ for each operator. The factor ${\rm R}^{(4)}$ is determined by supersymmetry 
\begin{equation}\label{R4}
{\rm R}^{(4)}=t_{12}^2t_{34}^2 x_{13}^4x_{24}^4(1-z\alpha)(1-\bar{z}\alpha)(1-z\bar{\alpha})(1-\bar{z}\bar{\alpha})\;,
\end{equation}
and doubles the $\mathcal{N}=2$ factor (\ref{R2}). Here $t_{ij}=t_i\cdot t_j$, and 
\begin{equation}\label{sigmatau}
\alpha\bar{\alpha}=\frac{t_{13}t_{24}}{t_{12}t_{34}}=\sigma\;,\quad (1-\alpha)(1-\bar{\alpha})=\frac{t_{14}t_{23}}{t_{12}t_{34}}=\tau\;.
\end{equation}

The full correlator $G_{k_1k_2k_3k_4}^{\mathcal{N}=4}$ gives rise to the full amplitude $\mathcal{M}_{k_1k_2k_3k_4}^{\mathcal{N}=4}$ via (\ref{MellinG}). The $\mathcal{N}=4$ reduced amplitude is similarly given by
\begin{equation}
\nonumber H_{k_1k_2k_3k_4}^{\mathcal{N}=4}\!=\!\!\!\int_{-i\infty}^{i\infty}\!\!\!\![dsdt]{\rm K}(x_{ij}^2;s,t,\tilde{u}) \widetilde{\mathcal{M}}_{k_1k_2k_3k_4}^{\mathcal{N}=4}\Gamma_{\{k_i\}}(s,t,\tilde{u})\,.
\end{equation}
But note here that the shift in $\tilde{u}$ is by $4$, {\it i.e.}, $\tilde{u}=u-4$. The greater shift is due to the higher conformal weights of ${\rm R}^{(4)}$. Bose symmetry again permutes $s$, $t$, $u$ in $\mathcal{M}_{k_1k_2k_3k_4}^{\mathcal{N}=4}$, and $s$, $t$, $\tilde{u}$ in $\widetilde{\mathcal{M}}_{k_1k_2k_3k_4}^{\mathcal{N}=4}$. At AdS tree level, the protected part again does not contributes to the Mellin amplitude \cite{Rastelli:2016nze,Rastelli:2017udc}. Therefore the full amplitudes are determined by the reduced amplitudes via  
\begin{equation}
\mathcal{M}_{k_1k_2k_3k_4}^{\mathcal{N}=4}=\mathbb{R}^{(4)}\circ \widetilde{\mathcal{M}}_{k_1k_2k_3k_4}^{\mathcal{N}=4}
\end{equation}
where we have promoted ${\rm R}^{(4)}$ into a difference operator $\mathbb{R}^{(4)}$ \cite{Rastelli:2016nze,Rastelli:2017udc}. The action of each monomial $U^mV^n$ in ${\rm R}^{(4)}/((t_{12})^2(t_{34})^2 x_{13}^4x_{24}^4)$ is given by (\ref{Oaction}) with $\mathcal{N}=4$. 

\section{Super gluon amplitudes}
We are now ready to discuss holographic correlators in specific theories. We start with super gluons in AdS$_5$ preserving $\mathcal{N}=2$ superconformal symmetry, which can be realized as D3 branes probing F theory singularities \cite{Fayyazuddin:1998fb,Aharony:1998xz}, or as $\mathcal{N}=4$ SYM with probe flavor D7 branes \cite{Karch:2002sh}. In both case, there is an AdS$_5$$\times$S$^3$ subspace in the holographic description, on which live localized degrees of freedom transforming in the adjoint representation of a color group $G_F$. These degrees of freedom form a vector multiplet, and its Kaluza-Klein reduction gives infinite towers of $\frac{1}{2}$-BPS superconformal multiplets. We refer to the $\frac{1}{2}$-BPS superprimaries as super gluons. At large central charge, gravity decouples and one has only a spin-1 gauge theory. Note S$^3$ has isometry $SO(4)=SU(2)_R\times SU(2)_L$. The first factor is identified with the $\mathcal{N}=2$ R-symmetry group, while the second $SU(2)_L$ is a global symmetry suppressed in the above discussion. The  operator $\mathcal{O}_k$ has spin $\frac{k-2}{2}$ under $SU(2)_L$ \cite{Aharony:1998xz}. We can similarly contract the indices with $k-2$ $SU(2)_L$ spinors $\bar{v}^{\bar{a}}$, $\bar{a}=1,2$. In reduced correlators, $v$ and $\bar{v}$ further recombine into null vectors of $SO(4)$ via Pauli matrices, and appear only as polynomials of $t_{ij}$  \cite{Alday:2021odx}
\begin{equation}
t^{r'}=\sigma^{r'}_{a\bar{a}}v^a\bar{v}^{\bar{a}}\;,\;\;\;{r'}=1,\ldots,4\;,\quad t\cdot t=0\;.
\end{equation}

To write down the super gluons amplitudes, let us choose, without loss of generality, the ordering $k_1\leq k_2\leq k_3\leq k_4$, and distinguish two cases
\begin{equation}
\nonumber k_1+k_4\geq k_2+k_3\,(\text{case I})\,,\;\;k_1+k_4< k_2+k_3\;(\text{case II})\,.
\end{equation}
To measure the deviation from the equal weight case $k_i=\frac{\Sigma}{4}$, it is useful to introduce the following parameters
\begin{equation}\label{kappa}
\begin{split}
&\kappa_s=|k_3+k_4-k_1-k_2|,\,\kappa_t=|k_1+k_4-k_2-k_3|,\\
&\kappa_u=|k_1+k_3-k_2-k_4|\,.
\end{split}
\end{equation}
The reduced Mellin amplitudes are given by \cite{Alday:2021odx} \footnote{Here we have set the gauge coupling to a convenient value which does not affect the physics.}
\begin{equation}
\begin{split}
\nonumber &\widetilde{\mathcal{M}}_{k_1k_2k_3k_4}^{\mathcal{N}=2}=\!\!\!\!\!\! \sum_{\substack{i+j+k = \mathcal{E} -2 \\ 0 \leq i,j,k \leq \mathcal{E}-2}} \!\frac{\sigma^i \tau^j}{i!\,j!\,k!\, (\tfrac{2i+\kappa_u}{2})!\, (\tfrac{2j+\kappa_t}{2})!\, (\tfrac{2k+\kappa_s}{2})!}\\
&\times\bigg[\frac{\mathtt{n}_s^{i,j}\mathtt{c}_s}{s-s_M+2k}+\frac{\mathtt{n}_t^{i,j}\mathtt{c}_t}{t-t_M+2j}+\frac{\mathtt{n}_u^{i,j}\mathtt{c}_u}{\tilde{u}-u_M+2i}\bigg]\times {\rm I}(t_{ab})
\end{split}
\end{equation}
which has been rewritten to manifest Bose symmetry. Let us unpack this expression a bit. Here
\begin{equation}
\nonumber\mathcal{E}=\frac{k_1+k_2+k_3-k_4}{2}\,(\text{case I})\,,\quad\mathcal{E}=k_1\,(\text{case II})\,
\end{equation}
is the {\it extremality}, which determines the complexity of the amplitude. After extracting a factor in $t_{ab}$
\begin{equation}
{\rm I}(t_{ab})=t_{34}^{\frac{\kappa_s}{2}}t_{24}^{\frac{\kappa_u}{2}}(t_{12}t_{34})^{-\mathcal{E}+2}\times \left\{\begin{array}{cc}t_{14}^{\frac{\kappa_t}{2}} & (\text{case I}) \\t_{23}^{\frac{\kappa_t}{2}} & (\text{case II})\end{array}\right.,
\end{equation}
the reduced Mellin amplitudes are degree-$(\mathcal{E}-2)$ polynomials in $\sigma$ and $\tau$ defined in (\ref{sigmatau}). The color dependence is captured by the color factors
\begin{equation}
\nonumber \mathtt{c}_s=f^{I_1I_2J}f^{JI_3I_4},\;\mathtt{c}_t=f^{I_1I_4J}f^{JI_2I_3},\;\mathtt{c}_u=f^{I_1I_3J}f^{JI_2I_4}
\end{equation}
where $f^{IJK}$ are the structure constants of the color group $G_F$. Thanks to the Jacobi identity, they satisfy $\mathtt{c}_s+\mathtt{c}_t+\mathtt{c}_u=0$. The kinematic factors $\mathtt{n}_{s,t,u}^{i,j}$ are given by
\begin{eqnarray}
\nonumber\mathtt{n}_s^{i,j}&=&\frac{1}{t-t_M+2j}-\frac{1}{\tilde{u}-u_M+2i}\;,\\
\mathtt{n}_t^{i,j}&=&\frac{1}{\tilde{u}-u_M+2i}-\frac{1}{s-s_M+2k}\;,\\
\nonumber\mathtt{n}_u^{i,j}&=&\frac{1}{s-s_M+2k}-\frac{1}{t-t_M+2j}\;.
\end{eqnarray}
The non-locality of these expressions is only superficial, and should not raise any alarm. In fact, a similar phenomenon occurs in flat space \footnote{In flat-space $\mathcal{N}=4$ SYM, the four-point superamplitude at tree level can be similarly written as $\delta^4(Q)\delta^4(\tilde{Q})(\frac{\mathtt{c}_sn_s}{s}+\frac{\mathtt{c}_tn_t}{t}+\frac{\mathtt{c}_un_u}{u})$. $\delta^4(Q)\delta^4(\tilde{Q})$ is roughly identified with ${\rm R}^{(2)}$, and the numerators $n_s=\frac{1}{3}(\frac{1}{t}-\frac{1}{u})$, $n_t=\frac{1}{3}(\frac{1}{u}-\frac{1}{s})$, $n_u=\frac{1}{3}(\frac{1}{s}-\frac{1}{t})$ are similarly non-local. However, this non-locality is merely an artifact of manifesting all the supersymmetry. We thank H. Johansson for comments on this}.  Evidently, $\mathtt{n}_{s,t,u}^{i,j}$ obey
\begin{equation}\label{ckduality}
\mathtt{n}_s^{i,j}+\mathtt{n}_t^{i,j}+\mathtt{n}_u^{i,j}=0\;,
\end{equation}
which gives rise to a realization of the {\it color-kinematic duality} \cite{Bern:2008qj} in AdS. In contrast to the duality pointed out in \cite{Alday:2021odx}, this new realization has the {\it same} form for both massless ($k_i=2$) and massive ($k_i>2$) super gluons. Finally, the remaining parameters are given by
\begin{eqnarray}
\nonumber s_M&=&\min\{k_1+k_2,k_3+k_4\}-2\;,\\
t_M&=&\min\{k_1+k_4,k_2+k_3\}-2\;,\\
\nonumber u_M&=&\min\{k_1+k_3,k_2+k_4\}-2\;.
\end{eqnarray}

\section{Super graviton amplitudes}
Let us now take a further step with the color-kinematic duality (\ref{ckduality}), and replace color factors $\mathtt{c}_{s,t,u}$ in each monomial $\sigma^i\tau^j$ by  kinematic factors $\mathtt{n}^{i,j}_{s,t,u}$. The result is
\begin{equation}
\begin{split}
\nonumber &\widetilde{\mathcal{M}}_{k_1k_2k_3k_4}^{\mathcal{N}=2\otimes\mathcal{N}=2}=\!\!\!\!\!\! \sum_{\substack{i+j+k = \mathcal{E} -2 \\ 0 \leq i,j,k \leq \mathcal{E}-2}} \!\frac{\sigma^i \tau^j}{i!\,j!\,k!\, (\tfrac{2i+\kappa_u}{2})!\, (\tfrac{2j+\kappa_t}{2})!\, (\tfrac{2k+\kappa_s}{2})!}\\
&\;\quad\quad\quad\quad\times \frac{-9\,{\rm I}(t_{ab})}{(s-s_M+2k)(t-t_M+2j)(\tilde{u}-u_M+2i)}\;.
\end{split}
\end{equation}
To interpret it as $\mathcal{N}=4$ reduced amplitudes, we need to {\it replace} the $\tilde{u}$ variable with the $\mathcal{N}=4$ one, as required by Bose symmetry of $\widetilde{\mathcal{M}}_{k_1k_2k_3k_4}^{\mathcal{N}=4}$. Furthermore, we replace the $SO(4)$ vectors $t^{r'}$ by $SO(6)$ null vectors \footnote{Note that the dimensionality of the SO group is invisible in scalar products $t_{ab}$.}. Remarkably, it gives all the super graviton reduced Mellin amplitudes of IIB supergravity on AdS$_5$$\times$S$^5$ \cite{Rastelli:2016nze,Rastelli:2017udc}
\begin{equation}
\widetilde{\mathcal{M}}_{k_1k_2k_3k_4}^{\mathcal{N}=4}=\sqrt{k_1k_2k_3k_4}\times \widetilde{\mathcal{M}}_{k_1k_2k_3k_4}^{\mathcal{N}=2\otimes\mathcal{N}=2}\;,
\end{equation}
up to an overall factor \footnote{The $k_i$-dependent normalization factor was fixed in \cite{Aprile:2018efk}. We have also set the Newton constant to a convenient value.}. This generalizes the {\it double copy relation} \cite{Bern:2010ue} into AdS space for four-point functions \footnote{In flat space the four-point superamplitude of $\mathcal{N}=8$ supergravity has the form $\delta^8(Q)\delta^8(\tilde{Q})(\frac{-1}{stu})$, where the doubled supercharge delta functions correspond to ${\rm R}^{(4)}$. Note the supergravity amplitude is also related to the flat space SYM amplitude in footnote \cite{Note8} by the double copy relation $\mathtt{c}_{s,t,u}\to n_{s,t,u}$}. In fact, redefining the super gravitons by $\mathcal{O}_k\to\mathcal{O}_k/\sqrt{k}$ gets rid of the normalization factor, and  gives the super graviton three-point functions also as the square of the super gluon ones \footnote{The new normalization makes the three-point function coefficients of super gravitons \cite{Lee:1998bxa} independent of $k_i$, {\it i.e.}, $C_{k_1k_2k_3}\sim 1$. The super gluons have exactly the same three-point functions \cite{Alday:2021odx}}.

\section{Bi-adjoint scalar amplitudes}
In flat space one can also replace kinematic factors by color factors, and obtains amplitudes of bi-adjoint scalars. We show that the same happens in AdS, and it serves as a nontrivial check. Note that in the above example the superconformal factor ${\rm R}^{(2)}$ was doubled to ${\rm R}^{(4)}$ ({\it c.f.}, (\ref{R2}) and (\ref{R4})). Going in the opposite direction, we expect ${\rm R}^{(0)}=1$, {\it i.e.}, the resulting theory has no supersymmetry. Moreover, since the internal spaces changed from S$^3$ to S$^5$, a reasonable guess is that this sequence starts with S$^1$, which will soon be confirmed. The symmetry groups are therefore $SO(\mathcal{N}+2)$, and we recall that operators in the reduced amplitudes transform in the rank-$(k_i-2)$ symmetric traceless representation.

Note that for $\mathcal{N}=0$, the null polarization vectors are two-component. Since we can rescale the null vectors, we are left with two inequivalent choices
\begin{equation}
t_{\pm}=\tfrac{1}{\sqrt{2}}(1,\pm i)\;.
\end{equation}
The dimension $k$ operator $\mathcal{O}^\pm_k\equiv \mathcal{O}_k(x,t_\pm)$ has $\pm (k-2)$ charges under $U(1)=SO(2)$, depending on the polarization chosen. Moreover, we assume the scalar interactions are only cubic. Then $U(1)$ charge conservation dictates that at least one of the $\kappa_s$, $\kappa_t$, $\kappa_u$ parameters in (\ref{kappa}) is zero. For the chosen ordering $k_1\leq k_2\leq k_3\leq k_4$, we must impose the condition $\kappa_t=0$. This leaves
\begin{equation}
\langle \mathcal{O}_{k_1}^+\mathcal{O}_{k_2}^-\mathcal{O}_{k_3}^-\mathcal{O}_{k_4}^+\rangle\;,\;\;{\rm or}\;\; \langle \mathcal{O}_{k_1}^-\mathcal{O}_{k_2}^+\mathcal{O}_{k_3}^+\mathcal{O}_{k_4}^-\rangle\;,
\end{equation}
which have identical amplitudes \footnote{They are the charge conjugation of each other.}. Noting
\begin{equation}
\sigma=1\;,\quad \tau=0\;, \quad {\rm I}(t_{ab})=1\;,
\end{equation}
and replacing $\mathtt{n}^{ij}_{s,t,u}$ with the color factors $\mathtt{c}'_{s,t,u}$ for another color group $G'_F$, we find 
\begin{equation}\label{Mbiadjoint}
\begin{split}
&\mathcal{M}_{k_1k_2k_3k_4}^{\mathcal{N}=0}=\!\!\!\!\!\! \sum_{\substack{i+k = \mathcal{E} -2 \\ 0 \leq i,k \leq \mathcal{E}-2}} \!\frac{-2\mathcal{N}_{k_1k_2k_3k_4}}{i!\,k!\, (\tfrac{2i+\kappa_u}{2})!\, (\tfrac{2k+\kappa_s}{2})!}\\
&\quad\quad\times\bigg[\frac{\mathtt{c}_s\mathtt{c}'_s}{s-s_M+2k}+\frac{\mathtt{c}_t\mathtt{c}'_t}{t-t_M}+\frac{\mathtt{c}_u\mathtt{c}'_u}{u-u_M+2i}\bigg]\;.
\end{split}
\end{equation}
We dropped the tildes because non-supersymmetric theories have only full amplitudes (\ref{MellinG}), and there is no shift in the $u$ variable. We also included a to-be-determined $k_i$-dependent normalization factor $-2\mathcal{N}_{k_1k_2k_3k_4}$ as in the supergravity case.  Remarkably, (\ref{Mbiadjoint}) can be rewritten as the sum of {\it three} AdS$_5$ scalar exchange diagrams
\begin{equation}\label{Mbiadjointb}
\mathcal{N}_{k_1k_2k_3k_4}\bigg(\frac{\mathtt{c}_s\mathtt{c}'_s}{p_s-1}\mathcal{S}^{(s)}_{p_s}+\frac{\mathtt{c}_t\mathtt{c}'_t}{p_t-1}\mathcal{S}^{(t)}_{p_t}+\frac{\mathtt{c}_u\mathtt{c}'_u}{p_u-1}\mathcal{S}^{(u)}_{p_u}\bigg)
\end{equation}
where $\mathcal{S}^{(s)}_{p}$ is the amplitude of exchanging a dimension-$p$ scalar in the s-channel (and similarly for the other two channels) \footnote{Here the normalization is such that the scalar operator $p$ has unit OPE coefficient.}
\begin{equation}
\begin{split}
\nonumber &\mathcal{S}^{(s)}_{p}=\!\!\sum_{m=0}^\infty \! \frac{-2(\frac{2+p-k_1-k_2}{2})_m(\frac{2+p-k_3-k_4}{2})_m}{(s-p-2m)m!(p-1)_m \Gamma[\frac{k_1+k_2-p}{2}]\Gamma[\frac{k_3+k_4-p}{2}]}\\
&\quad\quad\quad\;\; \times \frac{\Gamma[p]}{\Gamma[\frac{k_1-k_2+p}{2}]\Gamma[\frac{k_2-k_1+p}{2}]\Gamma[\frac{k_3-k_4+p}{2}]\Gamma[\frac{k_4-k_3+p}{2}]}\;.
\end{split}
\end{equation}
Moreover, the weights $p_{s,t,u}$ are precisely those selected by  $U(1)$ charge conservation
\begin{equation}
\nonumber p_s=k_2-k_2+2\;,\quad p_t=k_1+k_4-2\;,\quad p_u=k_3-k_1+2\;.
\end{equation}
Note that (\ref{Mbiadjoint}) is equivalent to (\ref{Mbiadjointb}) is highly nontrivial, and {\it a priori} does not need to happen. We can further fix the normalization $\mathcal{N}_{k_1k_2k_3k_4}$ by noting  $\mathcal{N}_{k_1k_2k_3k_4}/(p_s-1)$ {\it etc}, have the interpretation of products of three-point function coefficients $C_{k_1k_2p_s}C_{k_3k_4p_s}$. The solution, up to a $k_i$-independent overall factor, is  
\begin{equation}\label{Cijk}
C_{k_1k_2k_3}=\tfrac{1}{\sqrt{(k_1-1)(k_2-1)(k_3-1)}}\;.
\end{equation} 

Finally, we confirm by direct calculation that the theory is conformally coupled scalars on AdS$_5$$\times$S$^1$. The conformal mass on this manifold is $M^2_{\rm conf}=-4$ \footnote{We have set $R_{\rm AdS}=R_{\rm S}=1$.}. Decomposing the scalar field $\phi$ into S$^1$ modes $\phi(z,\tau)=\sum_{n=-\infty}^\infty \varphi_n (z) e^{in\tau}$, we find each mode has mass $M_n^2=n^2-4$. This translates into a conformal dimension $|n|+2$, agreeing with our charge-dimension relation $n=\pm (k-2)$. We can further check  three-point functions. A cubic vertex $\phi^3$ in AdS$_5$$\times$S$^1$ gives rise to infinitely many AdS$_5$ cubic vertices $\sum \varphi_{n_1}\varphi_{n_2}\varphi_{n_3}$ where $\{n_i\}$ conserve the $U(1)$ charge. Using the result of \cite{Freedman:1998tz}, it is straightforward to show that three-point functions are precisely (\ref{Cijk}). Note that both $C_{k_1k_2k_3}$ and $\mathcal{N}_{k_1k_2k_3k_4}$ can be set to one by redefining $\mathcal{O}_k \to \sqrt{k-1}\mathcal{O}_k$. Then the double copy relation also holds for three-point functions.

\section{Discussions}
In this note we found an extension of the double copy relation in curved spacetimes which relates all tree-level four-point functions of AdS$_5$$\times$S$^5$ IIB supergravity, AdS$_5$$\times$S$^3$ SYM, and AdS$_5$$\times$S$^1$ bi-adjoint scalars. Although our result is supersymmetric, it has immediate implications on {\it bosonic} Einstein gravity and YM theory in AdS$_5$ with no internal factor. Thanks to supersymmetry, four-graviton and four-gluon amplitudes can be obtained from the reduced correlators of  $k_i=2$ super gravitons and super gluons by action of differential operators \cite{Korchemsky:2015ssa}. At tree level these spinning correlators are identical to the ones in bosonic theories because the exchanged fields are the same \footnote{This follows directly from the symmetry selection rules.}. Our result then indicates that the bosonic amplitudes should also be related by a double copy construction \footnote{The kinematic numerators in this case will likely admit a local form, but they will be gauge-dependent}, of which the details we will leave for a future work. Another interesting direction is to extend our results to higher points, although more data of holographic correlators is needed \footnote{The five-point function of AdS$_5$ massless super gravitons has been computed in \cite{Goncalves:2019znr}.}. While the focus here is AdS$_5$ amplitudes, double copy relations for other backgrounds are also worth exploring. In particular, the AdS$_7$ case \cite{Alday:2020lbp} admits similar definitions of reduced amplitudes \cite{Rastelli:2017ymc,Zhou:2017zaw}.  Finally, it would be  interesting to explore extensions at higher genus, where the relevant CFT techniques were developed in \cite{Aharony:2016dwx}.

\vspace{0.3cm}

\noindent{\bf Acknowledgements.} I thank Fernando Alday, Henrik Johansson and Lionel Mason for very helpful comments on the manuscript. This work is supported in part by the Simons Foundation Grant No. 488653.

\bibliography{refs}
\bibliographystyle{utphys}

\end{document}